\documentclass[english]{article}
\pdfoutput=1
\usepackage[latin9]{inputenc}
\usepackage{floatflt}
\usepackage{amsmath}
\usepackage{graphicx}
\usepackage{amssymb}
\usepackage{esint}
\usepackage[authoryear]{natbib}

\makeatletter
\newcommand{\lyxaddress}[1]{
\par {\raggedright #1
\vspace{1.4em}
\noindent\par}
}

\newcommand{\captionfonts}{\small}

\makeatletter  
\long\def\@makecaption#1#2{%
  \vskip\abovecaptionskip
  \sbox\@tempboxa{{\captionfonts #1: #2}}%
  \ifdim \wd\@tempboxa >\hsize
    {\captionfonts #1: #2\par}
  \else
    \hbox to\hsize{\hfil\box\@tempboxa\hfil}%
  \fi
  \vskip\belowcaptionskip}
\makeatother   

\usepackage{babel}
\makeatother

\begin{document}

\title{Introduction to dissipation and decoherence in\\
quantum systems}

\author{F. Marquardt$^{\text{1}}$ and A. Püttmann$^{\text{2}}$}

\maketitle

\lyxaddress{(1) Arnold Sommerfeld Center for Theoretical Physics, Center for
NanoScience, and Department of Physics, Ludwig-Maximilians Universität
München, Theresienstr. 37, 80333 Munich, Germany\\
(2) Ruhr-Universität Bochum, Fakultät für Mathematik, Universitätsstraße
150, 44780 Bochum, Germany}

\begin{abstract}
These lecture notes address an audience of physicists or mathematicians
who have been exposed to a first course in quantum mechanics. We start
with a brief discussion of the general {}``system-bath'' paradigm
of quantum dissipative systems, analyze in some detail the simplest
example of {}``pure dephasing'' of a two-level system, and review
the basic concept of the density matrix. We then treat the general
dissipative time-evolution, introducing completely positive maps,
their relation to entanglement theory, and their Kraus decomposition.
Restricting ourselves to Markovian evolution, we discuss the Lindblad
form of master equations. The notes conclude with an overview of topics
of current interest that go beyond Lindblad Markov master equations.\\
\emph{These notes were prepared for lectures delivered by F.~Marquardt
in October 2007 at the Langeoog workshop of the SFB/TR 12, {}``Symmetries
and Universality in Mesoscopic Systems''}
\end{abstract}

\section{Introduction}

The following general situation is of interest in many fields of quantum
physics, ranging from quantum optics to condensed matter physics:
A single quantum system interacts with a large reservoir, alternatively
called {}``bath'' or {}``environment''. Whenever the system is
driven out of equilibrium by external perturbations, this coupling
makes the system relax back to equilibrium. \newcommand{\prl}{Phys. Rev. Lett.}

\begin{floatingfigure}{0.4\columnwidth}%
\begin{raggedright}
\includegraphics[width=0.4\columnwidth]{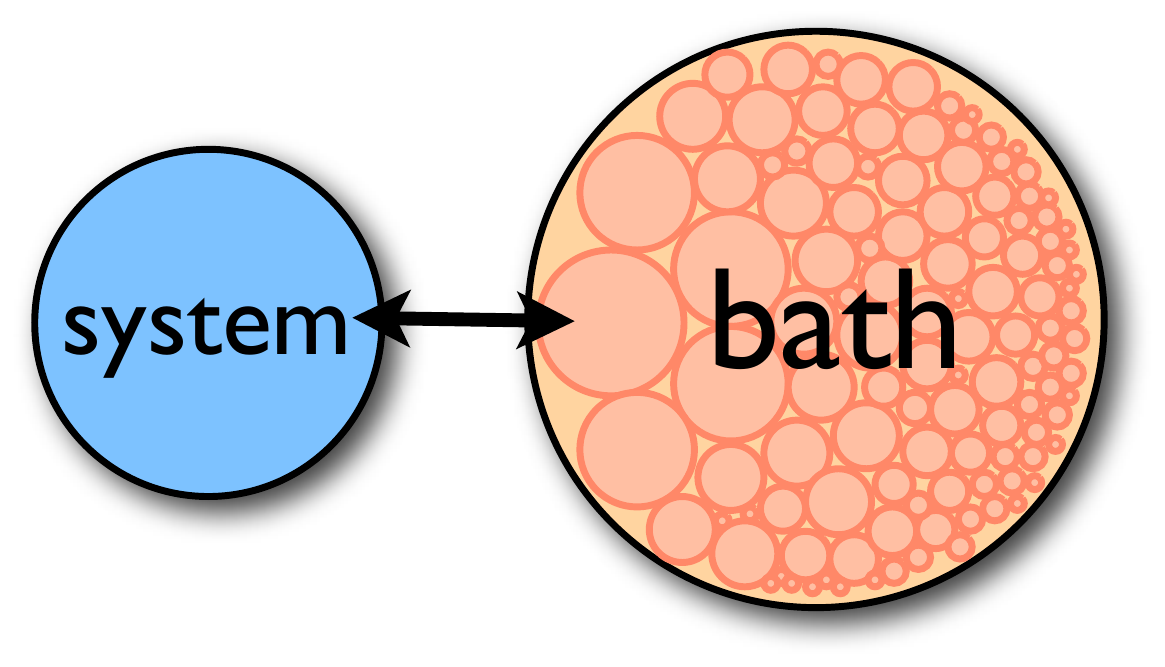}
\par\end{raggedright}

\caption{A quantum system interacting with the infinitely many degrees of freedom
of an environment.}
\end{floatingfigure}%
 The system itself might be a single atom, the spin degree of freedom
of an electron or a nucleus, a single particle moving through some
potential landscape (such as in a man-made interferometer in a semiconductor),
or even a many-particle system. In most of these cases the system
has only a few degrees of freedom, and sometimes it even has a finite-dimensional
Hilbert space (such as for the spin). In contrast, the bath needs
to have infinitely many degrees of freedom in order to generate truly
irreversible dynamics. It might be the electromagnetic field into
which the atom can radiate its energy, or the crystal lattice that
gets distorted by an electron moving along. 

In a classical setting, the bath introduces dissipation (friction),
as the system's energy can irreversibly be transferred to the bath.
By necessity, this also means that the system will experience a fluctuating
force. The balance of those two effects makes the system settle into
thermal equilibrium at a temperature set by the bath. In quantum dynamics,
there is yet another feature beyond these two: The system can display
coherent effects, i.e. interference phenomena in space or time. These
will be destroyed gradually by the coupling to the environment, an
effect known as {}``decoherence'' or {}``dephasing''.

In the following lecture notes, we will start with a simple example
that introduces the basic physics and requires only knowledge of elementary
quantum mechanics, applied to the two-level system. It will be used
in subsequent sections to illustrate the concepts. We then review
the description of {}``mixed'' (incoherent) quantum states by way
of the density matrix. Afterwards, we discuss the most general framework
for describing dissipative quantum dynamics: completely positive maps.
In the next step, we restrict ourselves to the important and simple
class of Markov dynamics and introduce the Lindblad master equation,
which is a workhorse of open systems dynamics in many branches of
physics. Finally, we outline a few topics of modern research that
go beyond this elementary tool.

\section{Example: Pure dephasing of a two-level system}

\newcommand{\ket}[1]{\left|#1\right\rangle }

\label{examplepuredeph}

\subsection{Effect of a classical stochastic process}

We consider a system described by a two-dimensional Hilbert space,
$\mathcal{H}=\mathbb{C}^{2}$. This is a two-level system, called
{}``qubit'' in some modern applications. Since the spin $1/2$ is
the most important physical realization, we denote the two basis states
as {}``spin up'' $\ket{\uparrow}$ and {}``spin down'' $\ket{\downarrow}$.
The vector describing the system's state contains the two complex
probability amplitudes for these basis states: 

\begin{equation}
\ket{\psi(t)}=\left(\begin{array}{c}
\psi_{\uparrow}(t)\\
\psi_{\downarrow}(t)\end{array}\right)\,.\end{equation}
In the absence of a magnetic field (i.e. in the absence of an energy
difference between the two spin states), $\psi_{\uparrow/\downarrow}(t)$
would be time-independent. We now look at the simplest possible model
for dephasing of this two-level system: The energy difference between
both basis states fluctuates in the course of time. For the moment,
we will not even keep the bath as a large quantum system in its own
right. Rather, we assume that the fluctuations it produces can be
treated as a classical stochastic process. This becomes a good approximation
for the limit of high temperatures. The time-dependent Hamiltonian
we thus want to study is

\begin{equation}
H(t)=\left(\begin{array}{cc}
\epsilon(t) & 0\\
0 & 0\end{array}\right)\,.\end{equation}
Here $\epsilon(t)$ is the stochastic process, which has the meaning
of a fluctuating magnetic field along the $z$-direction if we think
of the case of a spin. The Schrödinger equation $i\hbar\frac{d}{dt}\ket{\psi(t)}=H\ket{\psi(t)}$
can be solved directly. The components of the solution are

\begin{eqnarray}
\psi_{\downarrow}(t) & = & \psi_{\downarrow}(0)\\
\psi_{\uparrow}(t) & = & e^{i\varphi(t)}\psi_{\uparrow}(0),\end{eqnarray}
where the fluctuating phase is the integral over $\epsilon(t)$:

\begin{equation}
\varphi(t)=-\frac{1}{\hbar}\int_{0}^{t}\epsilon(t')\, dt'.\end{equation}
First of all, we observe that the probabilities $\left|\psi_{\uparrow/\downarrow}(t)\right|^{2}$
do not evolve, and only the relative phase between both states is
affected by the noise. This is the reason for speaking of {}``pure''
dephasing in this example. Next, we turn to look at observables that
are sensitive to the relative phase between the two basis states,
e.g. the operator $\sigma_{x}$ that is proportional to the spin's
component in the $x$-direction:

\begin{equation}
\sigma_{x}=\left(\begin{array}{cc}
0 & 1\\
1 & 0\end{array}\right).\end{equation}
Its quantum-mechanical expectation value, for a fixed realization
of the process $\epsilon(t)$, is given by

\begin{equation}
\left\langle \psi(t)\left|\sigma_{x}\right|\psi(t)\right\rangle =\psi_{\downarrow}^{*}(0)\psi_{\uparrow}(0)e^{i\varphi(t)}+{\rm c.c.}\end{equation}

\begin{figure}
\includegraphics[width=1\columnwidth]{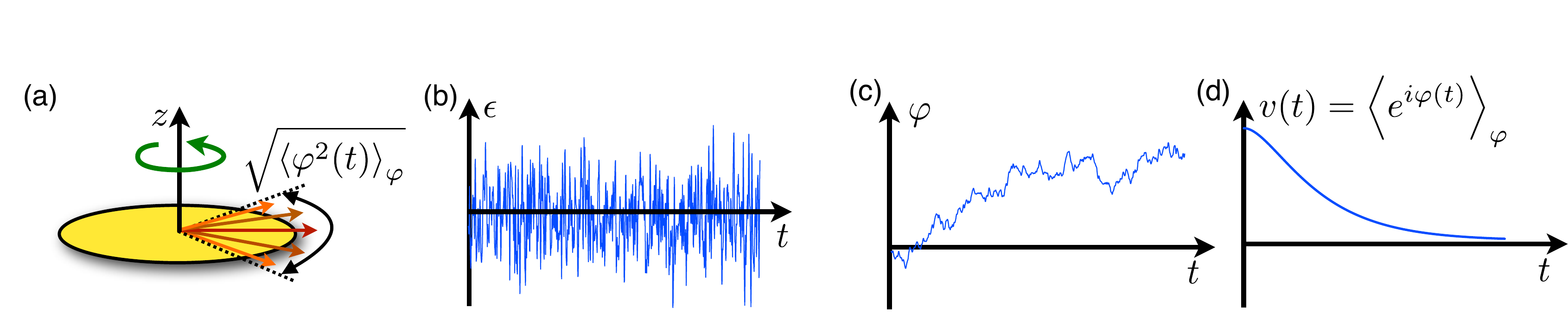}

\caption{(a) Pure dephasing can be viewed as arising from different, fluctuating
precession frequencies for an ensemble of spins; (b) Frequency fluctuations
as a stochastic process; (c) Resulting phase fluctuations; (d) Visibility}

\end{figure}
{[}Remarks on notation: We employ standard {}``bra-ket'' notation,
$\left\langle \psi\left|A\right|\phi\right\rangle \equiv\left\langle \psi|A\phi\right\rangle \equiv\psi^{\dagger}A\phi$,
where $\left\langle \psi\left|\psi\right.\right\rangle =1$. The complex
conjugated version of the first term on the right-hand-side is abbreviated
as ${\rm c.c.}$] 

In reality, it will be necessary to repeat the experiment many times
in order to obtain an estimate for the expectation value, by averaging
over the outcomes of $\sigma_{x}$ measurements (which yield either
$+1$ or $-1$ in each individual run). In general, each run will
correspond to a different realization of $\epsilon(t)$ as well. Therefore,
the result obtained by this averaging procedure actually also involves
taking the classical (stochastic) expectation value of the random
phase factor $e^{i\varphi}$. We will denote this average by $\left\langle \cdot\right\rangle _{\varphi}$
in order to distinguish it from the quantum mechanical expectation
value. Thus, the experimentalist will observe

\begin{equation}
\left\langle \left\langle \psi(t)|\sigma_{x}|\psi(t)\right\rangle \right\rangle _{\varphi}=\psi_{\downarrow}^{*}(0)\psi_{\uparrow}(0)\left\langle e^{i\varphi(t)}\right\rangle _{\varphi}+{\rm c.c.}\end{equation}
Apparently, the factor 

\begin{equation}
v(t)\equiv\left\langle e^{i\varphi(t)}\right\rangle _{\varphi}\end{equation}
describes the modification of the {}``interference term'' $\psi_{\downarrow}^{*}(0)\psi_{\uparrow}(0)$
due to the noise. The same factor would occur if we were to evaluate,
say, the expectation value of $\sigma_{y}$. We will here call this
factor {}``visibility'', as it has to do with the contrast or visibility
of interference experiments that rely on the coherence between the
two basis states. In particular, the overall magnitude of the interference
term is suppressed by $|v(t)|\leq1$. This will usually decrease in
the course of time, though that is not guaranteed and will depend
on the properties of the process $\epsilon(t)$. 

If $\epsilon$ is a Gaussian stochastic process (of zero mean), and
therefore $\varphi(t)$ is a Gaussian random variable with $\left\langle \varphi(t)\right\rangle =0$,
we can express the visibility explicitly in terms of the variance
of the phase:

\begin{equation}
v(t)=e^{-\frac{1}{2}\left\langle \varphi(t)^{2}\right\rangle _{\varphi}}.\end{equation}

\subsection{Several ways to obtain the same mixed state}

If we are only concerned with the system and the expectation values
of any of its observables (i.e. operators acting on the system alone),
there are indeed many more situations that will lead to the same predictions.
Let us enumerate them:

(a) \emph{The influence of a classical stochastic process}, with the
subsequent average, has been treated above.

(b) \emph{Interaction with a bath in a random initial state: }Suppose
we are considering a large Hilbert space, consisting of system and
bath: $\mathcal{H}_{SB}=\mathcal{H}_{S}\otimes\mathcal{H}_{B}$. The
product states of the form $\ket{\psi_{S}}\otimes\ket{\chi_{B}}\in\mathcal{H}_{SB}$
(where $\ket{\psi_{S}}\in\mathcal{H}_{S}$ and $\ket{\chi_{B}}\in\mathcal{H}_{B}$)
will be denoted as $\ket{\psi_{S}}\ket{\chi_{B}}$ for brevity, where
it is understood that the first part refers to the system, and the
second to the bath.

In order to describe the interaction between system and bath we might
either write down the full Hamiltonian, or else (more conveniently
for our purposes here) state the action of the full unitary time-evolution
operator $U_{SB}(t)$ that maps an initial state in $\mathcal{H}_{SB}$
onto a final state at time $t$. Given an orthonormal basis of states
$\ket{\chi_{j}}$ in $\mathcal{H}_{B}$, we postulate that the time-evolution
simply results in a phase factor that depends on the state, in the
form: 

\begin{eqnarray}
U_{SB}(t)\ket{\uparrow}\ket{\chi_{j}} & = & e^{i\varphi_{j}(t)}\ket{\uparrow}\ket{\chi_{j}}\label{eq:USB1}\\
U_{SB}(t)\ket{\downarrow}\ket{\chi_{j}} & = & \ket{\downarrow}\ket{\chi_{j}}\,.\label{eq:USB2}\end{eqnarray}
This defines $U_{SB}$. Now suppose furthermore that the initial state
of the bath is picked out of the $\ket{\chi_{j}}$ at random, each
of them occuring with a probability $w_{j}$ (where $\sum_{j}w_{j}=1$).
This means that the full initial state is $\ket{\psi_{SB}^{(j)}(0)}=\ket{\psi(0)}\ket{\chi_{j}}$
with probability $w_{j}$. Such a situation actually arises if the
bath is at a finite temperature, where different energy eigenstates
occur with classical probabilities fixed by the Boltzmann distribution. 

We now want to calculate the observed average value of $\sigma_{x}$
at time $t$. To this end, we must first evaluate the quantum-mechanical
expectation value of the product operator $\left(\sigma_{x}\right)_{S}\otimes\left(\mathbf{1}\right)_{B}$,
with respect to the state $\ket{\psi_{SB}^{(j)}(t)}=U_{SB}(t)\ket{\psi_{SB}^{(j)}(0)}$,
and then average according to the weights $w_{j}$. We write the operator
as $\sigma_{x}$ for short, it being understood that $\sigma_{x}$
acts only on the system and leaves the bath's state untouched. Straightforward
calculation using the rules (\ref{eq:USB1}) and (\ref{eq:USB2})
shows that we obtain exactly the same result as before:

\begin{equation}
\sum_{j}w_{j}\left\langle \psi_{SB}^{(j)}(t)\left|\sigma_{x}\right|\psi_{SB}^{(j)}(t)\right\rangle =\psi_{\downarrow}^{*}(0)\psi_{\uparrow}(0)\left\langle e^{i\varphi(t)}\right\rangle _{\varphi}+{\rm c.c.},\end{equation}
if we identify $\left\langle e^{i\varphi(t)}\right\rangle _{\varphi}$
with $\sum_{j}w_{j}e^{i\varphi_{j}}$ {[}the extension to a continuous
probability distribution is obvious]. Moreover, the results are the
same for \emph{any} system operator whose expectation value we care
to evaluate. In this sense, the incoherent ({}``mixed''; see below
for definition) system states produced according to (a) and (b) are
indistinguishable.

(c) \emph{{}``Quantum randomness''}: Instead of postulating classical
randomness in the choice of the initial bath state, we could as well
have postulated that it is a superposition of the basis states,

\begin{equation}
\ket{\chi(0)}=\sum_{j}\sqrt{w_{j}}\ket{\chi_{j}},\end{equation}
which is normalized since $\sum_{j}\left(\sqrt{w_{j}}\right)^{2}=1$.
Under the action of $U_{SB}(t)$, the initial state

\begin{equation}
\ket{\psi_{SB}(0)}=(\psi_{\uparrow}(0)\ket{\uparrow}+\psi_{\downarrow}(0)\ket{\downarrow})\otimes\ket{\chi(0)}\end{equation}
will evolve into

\begin{equation}
\ket{\psi_{SB}(t)}=\psi_{\uparrow}(0)\ket{\uparrow}\ket{\chi_{\uparrow}(t)}+\psi_{\downarrow}(0)\ket{\downarrow}\ket{\chi_{\downarrow}(t)}.\end{equation}
Here we have defined the two bath states $\ket{\chi_{\uparrow}(t)}$
and $\ket{\chi_{\downarrow}(t)}$ that have evolved out of $\ket{\chi(0)}$
under the influence of the system being in state $\ket{\uparrow}$
or $\ket{\downarrow}$, respectively:

\begin{eqnarray}
\ket{\chi_{\uparrow}(t)} & = & \sum_{j}\sqrt{w_{j}}e^{i\varphi_{j}(t)}\ket{\chi_{j}}\\
\ket{\chi_{\downarrow}(t)} & = & \sum_{j}\sqrt{w_{j}}\ket{\chi_{j}}=\ket{\chi(0)}.\end{eqnarray}
Again, when calculating the expectation value of $\sigma_{x}$, we
obtain the same result as above. However, in addition we realize that
the visibility $v(t)$ can now be expressed as the overlap of the
two bath states $\ket{\chi_{\uparrow/\downarrow}(t)}$:

\begin{equation}
\left\langle e^{i\varphi(t)}\right\rangle _{\varphi}=\sum_{j}w_{j}e^{i\varphi_{j}(t)}=\left\langle \chi_{\downarrow}(t)\left|\chi_{\uparrow}(t)\right.\right\rangle .\label{eq:overlap}\end{equation}
This is a result that is much more general than the present example
would suggest. The two states are sometimes called {}``pointer states'',
referring to the quantum theory of measurement. As it were, the environment
can be thought of as measuring the state of the system, which means
that $\ket{\chi(0)}$ evolves into either $\ket{\chi_{\uparrow}(t)}$
or $\ket{\chi_{\downarrow}(t)}$, depending on whether the system
had been in the state $\ket{\uparrow}$ or $\ket{\downarrow}$. This
is reminiscent of the pointer of a measuring device pointing either
way depending on the signal it picks up. When the two pointer states
become orthogonal, this implies that perfect knowledge about the system
could be inferred (in principle) from the state of the bath. As a
consequence, interference effects in the system itself are completely
destroyed, analogous to the gedanken experiment of the {}``Heisenberg
microscope'' \citep{1927_Heisenberg_UncertaintyRelation}.

In the most general framework, one can set up a path-integral analysis,
where the bath's state would evolve according to the particular trajectory
of the system. The overlap (\ref{eq:overlap}) between two states
having evolved under the influence of two different system trajectories
is then called the {}``Feynman-Vernon influence functional'' \citep{1963_FeynmanVernon_InfluenceFunctional}.
This forms the starting point for path-integral evaluations of the
dissipative dynamics of some systems, such as the damped quantum harmonic
oscillator or tunneling decay under the influence of dissipation.
For further details see the book of Weiss \citep{2000_Weiss_QuantumDissipativeSystems}.

\section{The density matrix}

The examples of the previous subsection all yield the same results
for the expectation values of any system operator. It is therefore
useful to introduce a description of the system's state that makes
this equivalence explicit and which is general enough to deduce from
it any arbitrary expectation value. This is achieved by the density
matrix, whose properties we review in this section. It forms the main
object of study in the field of open quantum systems, where the goal
will usually be to calculate the time-evolution of the density matrix
of a system subject to a fluctuating environment. While originally
the word {}``state of a quantum system'' referred only to {}``wave
functions'', i.e. vectors in a Hilbert space, it is now extended
to include the {}``mixed states'' that have to be described by a
density matrix. Most readers will know this material and can skip
directly to subsection \ref{sub:densmatpuredeph}, which returns to
the example of pure dephasing.

\subsection{Classical uncertainty}

There are different ways in which the density matrix concept may arise.
First, let us assume there is some classical uncertainty about the
system's state, i.e. the state $\ket{\psi_{j}}\in\mathcal{H}_{S}$
occurs with probability $w_{j}$. As explained above, when trying
to predict the experimentally observed average value of an observable
$A$, we thus should first take quantum-mechanical expectation values
and then perform an additional, {}``classical'' average: 

\begin{equation}
\left\langle A\right\rangle \equiv\sum_{j}w_{j}\left\langle \psi_{j}\left|A\right|\psi_{j}\right\rangle \end{equation}
We can obtain the same result by introducing the density matrix as
a weighted sum over projectors $\ket{\psi_{j}}\left\langle \psi_{j}\right|$,

\begin{equation}
\rho\equiv\sum_{j}w_{j}\ket{\psi_{j}}\left\langle \psi_{j}\right|,\label{eq:densmatdef}\end{equation}
and taking the trace over $\rho A$:

\begin{equation}
\left\langle A\right\rangle ={\rm tr}(\rho A).\label{eq:avgA}\end{equation}
Here $\rho\in B(\mathcal{H})$, where $B(\mathcal{H})$ denotes the
set of bounded operators on the Hilbert space $\mathcal{H}$.

In general, this step (from the set of $\ket{\psi_{j}}$ to $\rho$)
involves a compression of information. Regardless of the number of
states $\ket{\psi_{j}}$ we started out with in the beginning (which
need not be orthogonal!), $\rho$ itself is an operator on $\mathcal{H}_{S}$,
represented by a ${\rm dim}\mathcal{H}_{S}\times{\rm dim}\mathcal{H}_{S}$
matrix in the case of a finite-dimensional Hilbert space.

\subsection{General properties}

The following properties follow directly from the definition (\ref{eq:densmatdef}):

\begin{eqnarray}
{\rm tr}\rho & = & 1\label{eq:norm}\\
\rho^{\dagger} & = & \rho\label{eq:herm}\\
\rho & \geq & 0\label{eq:pos}\end{eqnarray}

For a {}``pure'' state $\ket{\psi}$, the density matrix is a projector
onto that state: $\rho=\ket{\psi}\left\langle \psi\right|$. As a
consequence, $\rho^{2}=\rho$ and ${\rm tr}\rho^{2}={\rm tr}\rho=1$.
In general, the {}``purity'' ${\rm tr}\rho^{2}$ obeys ${\rm tr}\rho^{2}\leq1$.
All states that are not pure are called {}``mixed'', and their density
matrix has a rank larger than one:

\begin{equation}
{\rm tr}\rho^{2}<1\Leftrightarrow\rho\,{\rm is}\,{\rm mixed}.\end{equation}

When making a measurement that is able to distinguish the particular
state $\ket{\phi}$ from other, orthogonal states, we can phrase this
by saying that we measure the operator $\ket{\phi}\left\langle \phi\right|$,
which will yield the value $1$ only when we indeed find $\ket{\phi}$.
Thus, the probability of observing $\ket{\phi}$ is the expectation
value of its projector:

\begin{equation}
P_{\phi}={\rm tr}\left(\rho\ket{\phi}\left\langle \phi\right|\right)=\left\langle \phi\left|\rho\right|\phi\right\rangle .\end{equation}
In the special case where $\rho$ describes the pure state $\ket{\psi}$,
this correctly reduces to the well-known postulate of quantum mechanics,
$P_{\phi}=\left|\left\langle \phi\right.\left|\psi\right\rangle \right|^{2}$.

After diagonalizing $\rho$, its eigenvalues $p_{j}$ can therefore
be interpreted as the probabilities of finding the respective eigenvectors
in a measurement that distinguishes between those eigenvectors. The
properties $\rho\geq0$ and ${\rm tr}\rho=1$ are then seen to correspond
to the simple fact that these probabilities are non-negative and normalized.

\subsection{Reduced density matrix}

We now turn to a different setting in which the density matrix arises:
Consider the world made up of a system and a bath, as explained in
the introduction. For generality, we will describe the overall state
of the world by a density matrix $\rho^{SB}$ (understood in the sense
explained above). Suppose we are now only interested in evaluating
expectation values of system operators $A_{S}$. Adapting (\ref{eq:avgA}),
we have

\begin{equation}
\left\langle A_{S}\right\rangle =\left\langle A_{S}\otimes\mathbf{1}_{B}\right\rangle ={\rm tr}_{SB}\left(\rho^{SB}(A_{S}\otimes\mathbf{1}_{B})\right).\end{equation}
Here we have indicated that the trace is taken in the full Hilbert
space $\mathcal{H}_{SB}$. However, since the operator acts only on
the system alone, we can break this trace into two steps:

\begin{equation}
\left\langle A_{S}\right\rangle ={\rm tr}_{S}\left({\rm tr}_{B}(\rho^{SB})A_{S}\right)\equiv{\rm tr}_{S}\left(\rho^{S}A_{S}\right).\end{equation}
Here we have introduced the reduced density matrix of the system,
by taking the partial trace over the bath degrees of freedom:

\begin{equation}
\rho^{S}\equiv{\rm tr}_{B}\left(\rho^{SB}\right)\end{equation}
We can make this explicit by choosing a product basis in $\mathcal{H}_{SB}$.
Then the matrix elements of $\rho^{S}$ are given by

\begin{equation}
\rho_{i'i}^{S}=\sum_{j}\rho_{(i',j)(i,j)}^{SB}\,,\end{equation}
where $(i,j)$ refers to the basis state $\ket{\psi_{i}}\ket{\chi_{j}}$,
with $\ket{\psi_{i}}\in\mathcal{H}_{S}$ and $\ket{\chi_{j}}\in\mathcal{H}_{B}$.

It is straightforward to confirm that the general properties (\ref{eq:norm}),(\ref{eq:herm}),
and (\ref{eq:pos}), continue to hold for $\rho^{S}$.

\subsection{Example: Pure dephasing}

\label{sub:densmatpuredeph}For the example described in section \ref{examplepuredeph},
the density matrix is given by

\begin{equation}
\rho(t)=\left(\begin{array}{cc}
\left|\psi_{\uparrow}(0)\right|^{2} & \psi_{\uparrow}(0)\psi_{\downarrow}^{*}(0)v(t)\\
\psi_{\uparrow}^{*}(0)\psi_{\downarrow}(0)v^{*}(t) & \left|\psi_{\downarrow}(0)\right|^{2}\end{array}\right).\end{equation}
This is independent of the specific model, i.e. it holds regardless
of whether we think of a classical noise process or of the interaction
with a quantum-mechanical environment (where $\rho$ would be the
reduced system density matrix). Only the off-diagonal element $\rho_{\uparrow\downarrow}(t)=\left\langle \uparrow\left|\rho(t)\right|\downarrow\right\rangle =\psi_{\uparrow}(0)\psi_{\downarrow}^{*}(0)v(t)$
is affected by dephasing. The decay of $\left|v(t)\right|$ will reduce
the purity of the state {[}we abbreviate $p=\left|\psi_{\uparrow}(0)\right|^{2}$]:

\begin{equation}
{\rm tr}\left(\rho^{2}(t)\right)=1-2p(1-p)(1-|v(t)|^{2})\end{equation}
When starting from an equal-weight superposition, $p=1/2$, this tends
to $1/2$ if $v\rightarrow0$. In that case, one ends up in the fully
mixed state: $\rho=\mathbf{1}/2$.

Let us have a look at the simplest possible case for the decay of
$v$, postulating that it decays exponentially in time:

\begin{equation}
v(t)=\left\langle e^{i\varphi(t)}\right\rangle _{\varphi}=e^{-\Gamma_{\varphi}t}\,,\end{equation}
where $\Gamma_{\varphi}$ is called the {}``dephasing rate''. Such
a decay will occur when the underlying fluctuations of $\epsilon$
are of the {}``white noise'' type, and therefore $\varphi(t)$ undergoes
Brownian motion ($\varphi$ is proportional to a Wiener process).
Then $\left\langle \varphi^{2}(t)\right\rangle _{\varphi}\propto t$,
which leads to exponential decay according to $\left\langle \exp[i\varphi(t)]\right\rangle _{\varphi}=\exp[-\left\langle \varphi(t)^{2}\right\rangle _{\varphi}/2]=\exp[-\Gamma_{\varphi}t]$.
For this particular case, we can write down a simple first-order differential
equation for the elements of $\rho(t)$:

\begin{equation}
\dot{\rho}_{\uparrow\downarrow}(t)=-\Gamma_{\varphi}\rho_{\uparrow\downarrow}(t)\,\,\,\,\,\,\dot{\rho}_{\uparrow\uparrow}=\dot{\rho}_{\downarrow\downarrow}=0\,.\label{eq:puredephmeq}\end{equation}
This is the simplest example of a {}``Markov master equation''.
The general structure of such an equation, which we will discuss later,
is of the form

\begin{equation}
\dot{\rho}=L\rho.\end{equation}

However, it is not permissible to postulate an arbitrary operator
$L$ on phenomenological grounds, because that might turn out to violate
the basic properties of $\rho$. This can be seen clearly in the following
example: Suppose we want to describe the spontaneous decay of the
excited state of an atom by emission of a photon. The first reasonable
(and indeed correct) ansatz that comes to mind would be to postulate
an exponential decay of the probability $\rho_{\uparrow\uparrow}$
of finding the atom in the excited state: $\dot{\rho}_{\uparrow\uparrow}=-\Gamma\rho_{\uparrow\uparrow}$,
and consequently $\dot{\rho}_{\downarrow\downarrow}=-\dot{\rho}_{\uparrow\uparrow}=+\Gamma\rho_{\uparrow\uparrow}$,
to conserve probability. Suppose we were to assume that the off-diagonal
elements do not change ($\dot{\rho}_{\uparrow\downarrow}=\dot{\rho}_{\downarrow\uparrow}=0$),
which seems to be the simplest possible assumption. In the long-time
limit $t\rightarrow\infty$, this would lead to 

\begin{equation}
\rho(t\rightarrow\infty)=\left(\begin{array}{cc}
0 & \rho_{\uparrow\downarrow}(0)\\
\rho_{\downarrow\uparrow}(0) & 1\end{array}\right)\,{\rm (wrong)}.\end{equation}
This is not a positive semidefinite matrix if $\rho_{\uparrow\downarrow}(0)\neq0$.
In other words, such an ansatz would violate a basic requirement,
the positivity of probabilities. The rest of these lecture notes is
concerned with describing the correct general structure of the time
evolution of density matrices, which makes sure that such pitfalls
are avoided.

\section{Completely positive maps and Kraus operators}

\subsection{Complete positivity}

Let us consider the linear map that takes the initial density matrix
$\rho(0)$ to its value at time $t$,

\begin{eqnarray}
\Phi:B(\mathcal{H}) & \rightarrow & B(\mathcal{H})\nonumber \\
\rho(0) & \mapsto & \rho(t)\,.\label{eq:PhiDef}\end{eqnarray}
Later on we will consider the map $\Phi$ for different times, but
for now we suppress the corresponding subscript for brevity. 

Let us list the properties which we will require of $\Phi$, which
follow from the properties of $\rho$ that we want to be respected
by the time-evolution:

\begin{eqnarray}
{\rm {\rm tr}\Phi(\rho)} & = & {\rm tr}\rho\,\,\,\,{\rm (norm-conserving)}\label{eq:normconserving}\\
\Phi(\rho^{\dagger}) & = & \Phi(\rho)^{\dagger}\,\,\,\,{\rm (preserves\,\, hermiticity)}\label{eq:hermconserving}\\
\rho\geq0 & \Rightarrow & \Phi(\rho)\geq0\,\,\,\,{\rm ("\Phi\, is\, a\, positive\, map")}\label{eq:positivemap}\end{eqnarray}
It would seem that these three features are all that is needed to
have a permissible time-evolution. Surprisingly, that is not the case.
The third requirement, of $\Phi$ being a positive map, has to be
replaced by the stronger condition of $\Phi$ being {}``completely
positive''.

\emph{Definition}\textendash{} We call a linear map $\Phi:B(\mathcal{H})\rightarrow B(\mathcal{H})$
{}``\emph{completely positive}'' (CP) iff the following holds: Given
any other Hilbert space $\mathcal{H}'$, we consider the product space
$\tilde{\mathcal{H}}=\mathcal{H}\otimes\mathcal{H}'$ and construct
a linear map $\tilde{\Phi}:B(\tilde{\mathcal{H}})\rightarrow B(\tilde{\mathcal{H}})$
that derives from $\Phi$ by acting like $\Phi$ onto the original
Hilbert space $\mathcal{H}$ and not affecting the space $\mathcal{H}'$,
that is: $\tilde{\Phi}(\rho\otimes\rho')=\Phi(\rho)\otimes\rho'$
for $\rho\in B(\mathcal{H})$ and $\rho'\in B(\mathcal{H}')$. Then
$\tilde{\Phi}$ is a positive map.

\subsection{Example of a positive but not completely positive map (relation to
entanglement theory)}

At first sight, it is surprising that positivity of $\Phi$ itself
is not enough to guarantee the positivity of the enlarged map $\Phi$.
We will now look at the simplest example of a map that is positive
but not completely positive. This example is of interest in the theory
of entanglement.

\emph{Claim}\textendash{} $\Phi:\rho\mapsto\rho^{t}$ is positive
but not CP.

It is obvious that $\Phi$ is positive, as transposition does not
change the eigenvalues of a matrix. Now let us consider the enlarged
map $\tilde{\Phi}$, operating on a product basis in $\tilde{\mathcal{H}}=\mathcal{H}\otimes\mathcal{H}'$.
It induces a {}``partial transposition'' of a density matrix $\tilde{\rho}\in B(\tilde{\mathcal{H}})$:

\begin{equation}
[\tilde{\Phi}(\tilde{\rho})]_{(i',j')(i,j)}=\tilde{\rho}_{(i,j')(i',j)}\,,\end{equation}
where we observe the interchange of indices $i$ and $i'$ referring
to $\mathcal{H}$. In order to prove that $\tilde{\Phi}$ is not a
positive map, it is enough to find one example of a state $\tilde{\rho}\geq0$
(for one suitably chosen Hilbert space $\mathcal{H}'$) for which
the partial transposition fails to make $\tilde{\Phi}(\tilde{\rho})\geq0$.
It is clear that any product state, of the type $\tilde{\rho}=\rho\otimes\rho'$,
will not be sufficient for this purpose, since $\rho^{t}\geq0$ and
therefore $\tilde{\Phi}(\tilde{\rho})\geq0$ as well. More generally,
due to the linearity of $\tilde{\Phi}$, the same holds for any so-called
{}``\emph{separable}'' state which is a mixture of product states:

\begin{equation}
\tilde{\rho}=\sum_{j}w_{j}\rho^{(j)}\otimes\rho'^{(j)}\,.\end{equation}
Here $\rho^{(j)},\rho'^{(j)}$ are valid density matrices in $B(\mathcal{H})$
and $B(\mathcal{H}')$, respectively, and $w_{j}\geq0,\,\sum_{j}w_{j}=1$.
Any separable state will yield a positive semi-definite partial transpose:
$\tilde{\Phi}(\tilde{\rho})\geq0$. This is the important PPT ({}``positive
partial transpose'') criterion, a \emph{necessary} condition for
separability of a state $\tilde{\rho}$, discovered by Asher Peres
in 1996 \citep{1996_Peres_SeparabilityCriterion}. 

In order to find an example without a positive partial tranpose, we
thus have to consider non-separable, i.e. so-called {}``\emph{entangled}''
states. The simplest case is having both $\mathcal{H}$ and $\mathcal{H}'$
two-dimensional Hilbert spaces, and considering the pure state $\tilde{\rho}=\ket{\tilde{\psi}}\left\langle \tilde{\psi}\right|$,
with

\begin{equation}
\ket{\tilde{\psi}}=\frac{1}{\sqrt{2}}(\ket{\uparrow\uparrow}+\ket{\downarrow\downarrow}).\end{equation}

\[
\]
The partial transposition acts like $\tilde{\Phi}(\ket{\uparrow\uparrow}\left\langle \downarrow\downarrow\right|)=\ket{\downarrow\uparrow}\left\langle \uparrow\downarrow\right|$
(note the interchange in the first position, referring to $\mathcal{H}$),
and likewise on other combinations occuring in the projector $\tilde{\rho}$.
As a result, we find

\begin{equation}
\tilde{\Phi}(\tilde{\rho})=\frac{1}{2}\left(\begin{array}{cccc}
1 & 0 & 0 & 0\\
0 & 0 & 1 & 0\\
0 & 1 & 0 & 0\\
0 & 0 & 0 & 1\end{array}\right)\,,\end{equation}
where the matrix has been written down with respect to the basis \\
$\ket{\uparrow\uparrow},\ket{\uparrow\downarrow},\ket{\downarrow\uparrow},\ket{\downarrow\downarrow}$.
It has eigenvalues $+1/2$ (three-fold degenerate) and $-1/2$. A
a consequence, $\tilde{\Phi}$ is not positive, and $\Phi$ is not
a CP map.

As a side-note, we mention that PPT is even a necessary and \emph{sufficient
}criterion for separability if ${\rm dim}\mathcal{H}={\rm dim}\mathcal{H}'=2$
as in our example {[}and this even holds when the dimensions are 2
and 3, respectively]. One has to go to higher-dimensional Hilbert
spaces in order to find states that are not separable (i.e. entangled)
but still have a positive partial transpose.

Physically, the property of complete positivity ensures that one ends
up with a permissible state even if the dissipative system had been
entangled initially with another system. As a consequence, there is
no analogon to the concept of complete positivity for classical dissipative
systems, since classical physics does not know entangled states.

\subsection{Kraus decompositions of a CP map}

It turns out that all CP maps that fulfill the other properties mentioned
above can be decomposed in a simple way. We state the theorem, due
to Kraus \citep{1983_Kraus_StatesEffects}, without proof:

\emph{Theorem}\textendash{} Provided we are given a map $\Phi$ that
fulfills the properties (\ref{eq:normconserving}) and (\ref{eq:hermconserving}),
as well as complete positivity, there exists a set of {}``Kraus operators''
$K_{j}:B(\mathcal{H})\rightarrow B(\mathcal{H})$ that are normalized
in the sense $\sum_{j}K_{j}^{\dagger}K_{j}=\mathbf{1}$ and that can
be used to represent $\Phi(\rho)$:

\begin{equation}
\Phi(\rho)=\sum_{j}K_{j}\rho K_{j}^{\dagger}.\end{equation}
The converse also holds (i.e. the three properties follow from the
existence of such a decomposition). In the case of an infinite-dimensional
Hilbert space, the set of values for the index $j$ can be countably
infinite.

We note that this decomposition is not at all unique. For example,
multiplying $K_{j}$ by a phase factor $e^{i\theta_{j}}$ changes
nothing. More generally, a unitary matrix with matrix elements $U_{j'j}$
can be used to convert to a new set of Kraus operators that represent
the same map: $K'_{j'}=\sum_{j}K_{j}U_{jj'}$ {[}where the cardinality
of the two sets is assumed to be the same, adding Kraus operators
equal to zero if need be]. It can be shown \citep{1998_preskill_notes}
that any two equivalent Kraus decompositions are connected in this
way. We also note that for a finite-dimensional Hilbert space any
map can be represented using at most $({\rm dim}\mathcal{H})^{2}$
Kraus operators.

\subsection{Examples}

We now list a few examples for the applications of Kraus decompositions:

\begin{itemize}
\item Purely unitary time-evolution: $\ket{\psi(t)}=U(t)\ket{\psi(0)}$
leads to\\
$\rho(t)=U(t)\rho(0)U^{\dagger}(t)$, and $\rho(t)=\Phi(\rho(0))$
can therefore be represented by the single Kraus operator $K=U(t)$.
\item Random unitary evolution: If we pick some $U_{j}$ at random with
probability $w_{j}$, as in the example with pure dephasing by classical
noise, then the set $K_{j}=\sqrt{w_{j}}U_{j}(t)$ will yield the correct
$\rho(t)=\Phi(\rho(0))=\sum_{j}w_{j}U_{j}(t)\rho(0)U_{j}^{\dagger}(t)$,
and normalization follows from $\sum_{j}w_{j}=1$.
\item However, in the example of pure dephasing of a two-level system, we
can also choose a more economical decomposition. For example, for
$v(t)\in\mathbb{R}$, just two Kraus operators suffice: $K_{1}=\mathbf{1}\sqrt{(1+v)/2}$
and\\
$K_{2}=\sigma_{z}\sqrt{(1-v)/2}$. 
\item The $\sigma_{z}$ Kraus operator in the previous example can be thought
of as describing a {}``phase flip'' (changing the sign of the off-diagonal
element of the density matrix). In the same sense, $\sigma_{x}$ would
describe a {}``bit flip'' (turning $\ket{\uparrow}$ into $\ket{\downarrow}$
and vice versa), and $\sigma_{y}$ a combination of the two. 
\item In quantum information processing, two-level systems are viewed as
{}``quantum bits'' (qubits). After sending such a qubit through
a communication channel (where it can be subject to technical noise,
or even interact with, and get entangled with, some bath), its state
will have been changed by a map $\Phi$ that is characteristic for
this quantum channel. Thus a quantum channel can be described by giving
a set of Kraus operators.
\item As indicated by the two-level example from above, for a finite-dimensional
Hilbert space a finite number of Kraus operators will suffice (even
if the underlying physical description of the noise involved an infinite
number of possible different time evolutions $U_{j}$). This is very
important in the theory of quantum error correction. It means that,
contrary to first appearances, quantum computers are not as bad as
classical analog computers when it comes to error correction.
\item Kraus operators can be used to describe measurements: After an ideal
von-Neumann measurement that distinguishes between the orthogonal
states $\ket{\phi_{j}}$, the system ends up in one of those states
with probability $p_{j}={\rm tr}(\rho\ket{\phi_{j}}\left\langle \phi_{j}\right|)$.
However, from the point of view of someone who is not told the measurement
outcomes, the state after the measurement is described by the density
matrix $\rho'=\sum_{j}p_{j}\ket{\phi_{j}}\left\langle \phi_{j}\right|$,
and the map from $\rho$ to $\rho'$ can be described by the Kraus
operators $K_{j}=\ket{\phi_{j}}\left\langle \phi_{j}\right|$. The
more general case of measurements that reveal only partial information
(POVM: positive operator valued measurements) is described by arbitrary
Kraus operators that are not necessarily projectors, and the probability
of finding a particular value is then ${\rm tr}\left(\rho K_{j}^{\dagger}K_{j}\right)$.
\item For systems with high- (or infinite-) dimensional Hilbert spaces (like
those with a continuous degree of freedom, e.g. a harmonic oscillator),
the Kraus decomposition, though still possible in principle, becomes
less useful in practice due to the large number of Kraus operators.
\end{itemize}

\subsection{Construction of Kraus operators for system-bath interaction}

We now give an explicit construction of the Kraus operators for the
important example where the time-evolution of the reduced density
matrix is due to the interaction with an environment. Suppose that
at time $t=0$ the environment is in the initial state $\ket{\chi(0)}$,
uncorrelated with the arbitrary initial system state $\rho^{S}(0)$.
This means for the total (system+bath) density matrix: $\rho^{SB}(0)=\rho^{S}(0)\otimes\ket{\chi(0)}\left\langle \chi(0)\right|$.
In addition, we allow for an arbitrary unitary time-evolution $U_{SB}$
acting on the product Hilbert space $\mathcal{H}_{SB}$, taking the
full initial state to the final state at time $t$. Therefore, the
system's reduced density matrix at time $t$ is: 

\begin{eqnarray}
\rho(t) & = & {\rm tr}_{B}\left(U_{SB}\rho_{S}\otimes\ket{\chi(0)}\left\langle \chi(0)\right|U_{SB}^{\dagger}\right)\nonumber \\
 & = & \sum_{j}\underbrace{\left\langle \chi_{j}\left|U_{SB}\right|\chi(0)\right\rangle }_{\equiv K_{j}}\rho_{S}\left\langle \chi(0)\left|U_{SB}^{\dagger}\right|\chi_{j}\right\rangle \nonumber \\
 & = & \sum_{j}K_{j}\rho_{S}K_{j}^{\dagger}\end{eqnarray}
In the second line we have introduced the sum over a basis $\ket{\chi_{j}}$
of the bath Hilbert space $\mathcal{H}_{B}$ to perform the trace
over the bath. The resulting matrix elements of the full time-evolution
$U_{SB}$ with respect to the bath states define the operators $K_{j}$,
as indicated. These operators act on the system Hilbert space $\mathcal{H}_{S}$.
They are normalized, since

\begin{equation}
\sum_{j}K_{j}^{\dagger}K_{j}=\sum_{j}\left\langle \chi(0)\left|U_{SB}^{\dagger}\right|\chi_{j}\right\rangle \left\langle \chi_{j}\left|U_{SB}\right|\chi(0)\right\rangle =1.\end{equation}
As a consequence, we can identify them as the Kraus operators needed
to describe the map of $\rho(0)=\rho^{S}(0)$ to $\rho(t)$. If the
bath initially were in a mixed state, we would find $\rho(t)$ to
be the weighted average of expressions of this type, and consequently
the full set of Kraus operators would consist of the individual sets,
multiplied by the factors $\sqrt{w}_{j}$, where $w_{j}$ are the
weights. In this way, the time-evolution due to interaction with an
environment (starting from an uncorrelated state) can always be expressed
using Kraus operators, i.e. as a CP map. The idea of the general construction
of Kraus decompositions, presented in \citep{1998_preskill_notes},
makes use of this fact.

\section{Markov master equations of Lindblad form}

In the previous section, we have been dealing with the completely
arbitrary time-evolution of a density matrix. Let us now specialize
to evolutions of the Markov type, i.e. where the density matrix follows
an equation of the form $\dot{\rho}=L\rho$, with $L$ representing
a linear map. This is called Markovian since the evolution of the
state only depends on the present state itself, not on its history.
Such an equation is the direct quantum analogue of the evolution equation
for the probability density for the case of a classical stochastic
process of Markov type.

\subsection{Lindblad's theorem}

It is clear that the Markov property, when combined with the restrictions
discussed in the previous section, will lead to a particular form
of $L$. This problem has first been considered in its full generality
by Lindblad \citep{1976_Lindblad_MasterEquation} (1976). In writing
down the assumptions for his proof, which we will list below, Lindblad
considers the map in the Heisenberg picture, where the time-evolution
is applied to the observable $A$ instead of the density matrix $\rho$:

\begin{equation}
{\rm tr}[\rho(t)A]={\rm tr}[\Phi_{t}(\rho(0))A]={\rm tr}[\rho(0)\Phi_{t}^{H}(A)],\end{equation}
for any choice of initial density matrix $\rho(0)$.

Then a \emph{{}``completely positive dynamical semigroup}'' is defined
as follows: Let $\mathcal{A}$ be a $W^{*}$-algebra%
\footnote{A $W^{*}$-algebra (also called a von-Neumann algebra), is a {*}-algebra
of bounded operators on a Hilbert space that is closed in the weak
{*} topology and contains the identity operator. This essentially
means it is an algebra of operators $A$, which can be added and multiplied
with a scalar as usual (thus forming a vector space), have the usual
rules of non-commutative multiplication between operators, the possibility
of taking the adjoint (that is what the {*} refers to) with all the
features you would expect, and an operator norm that is compatible
with the adjoint operation: $\left\Vert A^{\dagger}A\right\Vert =\left\Vert A\right\Vert ^{2}$.
See books on functional analysis, such as \citep{1972_ReedSimon_FunctionalAnalysisBook}.%
} and $\Phi_{t}$ be a family of completely-positive maps of $\mathcal{A}$
into itself, depending on the real-valued time parameter $t\in\mathbb{R}_{0}^{+}$.
In addition, we require the following properties:

\begin{eqnarray}
\Phi_{t}^{H}(\mathbf{1}) & = & \mathbf{1}\\
\Phi_{s}^{H}\cdot\Phi_{t}^{H} & = & \Phi_{s+t}^{H}\,\,\,(\forall s,t\geq0)\\
\lim_{t\rightarrow0}\left\Vert \Phi_{t}^{H}-\mathbf{1}\right\Vert  & = & 0\\
\Phi_{t}^{H} &  & {\rm is\, normal\,(ultraweakly\, continuous)}\end{eqnarray}
The first line guarantees conservation of the trace of the density
matrix. The second line is the central semi-group property, which
will guarantee Markov dynamics and represents a strong restriction
on the allowed physical situations. It tends to work as a good approximation
in situations where the correlation time of the fluctuations characterizing
the environment is short in comparison with other time scales, such
as those set by decay rates and oscillation periods. Under these conditions,
Lindblad proved the following:

The action of $\Phi_{t}$ can be expressed in the form of a Markov
master equation, where $\rho(t)=\Phi_{t}(\rho(0))$ fulfills the differential
equation 

\begin{equation}
\dot{\rho}(t)=L\rho(t),\end{equation}
with the Liouvillian operator of {}``Lindblad form'':

\begin{equation}
L\rho=-\frac{i}{\hbar}\left[H,\rho\right]+\sum_{j}(R_{j}\rho R_{j}^{\dagger}-\frac{1}{2}R_{j}^{\dagger}R_{j}\rho-\frac{1}{2}\rho R_{j}^{\dagger}R_{j})\,.\label{eq:Lindblad}\end{equation}
Here the first part, involving some hermitian Hamiltonian $H$, describes
the standard unitary time-evolution of $\rho$ (possibly with renormalized
matrix elements of $H$ due to the presence of the environment, i.e.
$H\neq H_{0}$, where $H_{0}$ would be the intrinsic Hamiltonian).
The dissipative dynamics is generated by the second term, with the
relaxation (or Lindblad) operators $R_{j}$ that do not have to fulfill
any special constraint (unlike the normalized Kraus operators, from
which they derive). We note that, just as for the Kraus operators,
the choice of $R_{j}$ is not unique, and even the separation into
a unitary and a dissipative part is not unique either.

\subsection{Obtaining the Lindblad form from the Kraus decomposition}

We now give the main ideas of the derivation, building on the Kraus
decomposition, without attempting mathematical rigour.

The density matrix at a small time $\delta t$ deviates from $\rho(t=0)$
only to first order in $\delta t$, due to continuity and the Markov
structure. In addition, it can be written in the form of a Kraus decomposition:

\begin{equation}
\rho(\delta t)=\rho(0)+\mathcal{O}(\delta t)=\sum_{j}K_{j}\rho(0)K_{j}^{\dagger},\label{eq:KrausDecompLindblad}\end{equation}
where $\sum_{j}K_{j}^{\dagger}K_{j}=\mathbf{1}$. In order to satisfy
this structure, we need one $ $Kraus operator close to unity, which
we will write as

\begin{equation}
K_{0}=\mathbf{1}+(-\frac{i}{\hbar}H+A)\delta t+\ldots\label{eq:K0}\end{equation}
The second term has been decomposed into {}``real'' and {}``imaginary''
parts, with two hermitean operators $H^{\dagger}=H$ and $A^{\dagger}=A$.
All the other Kraus operators $K_{j}$ (with $j\geq1$) must be of
order $\mathcal{O}(\sqrt{\delta t})$, to obtain the desired expression
(\ref{eq:KrausDecompLindblad}). We can now obtain $A$ from the normalization
requirement:

\begin{equation}
\mathbf{1}=\sum_{j\geq0}K_{j}^{\dagger}K_{j}=\mathbf{1}+2\delta tA+\sum_{j\geq1}K_{j}^{\dagger}K_{j},\end{equation}
which yields

\begin{equation}
A=-\frac{1}{2\delta t}\sum_{j\geq1}K_{j}^{\dagger}K_{j},\end{equation}
that is an expression of $\mathcal{O}(\delta t^{0})$. Inserting this
result first into (\ref{eq:K0}) and then into the decomposition (\ref{eq:KrausDecompLindblad}),
we obtain the Lindblad structure (\ref{eq:Lindblad}), by identifying
the relaxation operators as

\begin{equation}
R_{j}=\lim_{\delta t\rightarrow0}\frac{K_{j}}{\sqrt{\delta t}}\,\,\,(j\geq1).\end{equation}

\subsection{Examples}

We now list a few examples for the relaxation operators occuring for
different dissipative processes. In the present lecture notes, we
will not address the techniques used for obtaining microscopic derivations
of these operators and the rates occuring in there, which depend on
the specific type of environment and the assumed coupling. In general,
however, the relaxation operators are of a simple form, describing
the operator that induces the dissipative transition, multiplied with
the square root of the corresponding rate.

\begin{itemize}
\item Pure dephasing of a two-level system, with exponential decay of the
visibility at a rate $\Gamma_{\varphi}$: $R=\sqrt{\frac{\Gamma_{\varphi}}{2}}\sigma_{z}$
yields the Markov master equation (\ref{eq:puredephmeq}).
\item Exponential relaxation from the excited state to the ground state,
at a rate $\Gamma$: $R=\sqrt{\Gamma}\sigma_{-}$, where $\sigma_{-}=\ket{\downarrow}\left\langle \uparrow\right|$.
This yields the proper decay of the probabilities, $\dot{\rho}_{\uparrow\uparrow}=-\Gamma\rho_{\uparrow\uparrow}$
and $\dot{\rho}_{\downarrow\downarrow}=-\dot{\rho}_{\uparrow\uparrow}=+\Gamma\rho_{\uparrow\uparrow}$,
but it also gives non-trivial information on the decay of the off-diagonal
elements: \begin{equation}
\dot{\rho}_{\uparrow\downarrow}=-\frac{i}{\hbar}(\epsilon_{\uparrow}-\epsilon_{\downarrow})\rho_{\uparrow\downarrow}-\frac{\Gamma}{2}\rho_{\uparrow\downarrow}\,.\end{equation}
The first term results from the unitary evolution (i.e. from $H=\epsilon_{\uparrow}\ket{\uparrow}\left\langle \uparrow\right|+\epsilon_{\downarrow}\ket{\downarrow}\left\langle \downarrow\right|$),
whereas the second term describes dephasing at a rate $\Gamma_{\varphi}^{{\rm eff}}=\Gamma/2$
that is exactly half the total decay rate. If pure dephasing is present
in addition, one just needs to introduce a second relaxation operator,
as given above, and the rates would add: $\Gamma_{\varphi}^{{\rm total}}=\Gamma_{\varphi}+\Gamma/2$.
In the context of nuclear magnetic resonance or qubit physics, the
fact that $\Gamma_{\varphi}^{{\rm total}}\geq\Gamma/2$ is often expressed
as an inequality for the corresponding time-scales (inverses of the
rates): $T_{2}\leq2T_{1}$, where $T_{2}=(\Gamma_{\varphi}^{{\rm total}})^{-1}$
and $T_{1}=\Gamma^{-1}$. We have encountered this inequality here
ultimately as a consequence of the requirement that the time-evolution
preserve positivity of $\rho$ {[}which any decay rate smaller than
$\Gamma/2$ for the off-diagonal elements would not ensure].
\item If the bath is at finite temperature, the two-level system might also
absorb a quantum of energy, i.e. become thermally excited. This is
described by $R=\sqrt{\Gamma_{{\rm up}}}\sigma_{+}$, where $\sigma_{+}=\sigma_{-}^{\dagger}=\ket{\uparrow}\left\langle \downarrow\right|$.
\item Taking into account all the three processes discussed up to now leads
to the so-called {}``Bloch equations'' first invented to describe
the dissipation and decoherence of systems such as atoms or spins.
\item Damping of a harmonic oscillator (due to a coupling to the bath that
is linear in the coordinate of the oscillator) is described by $R=\sqrt{\Gamma}a$.
Here $\Gamma$ is the damping rate (which can be observed in the linear
response of the system), and $a$ is the annihilation operator that
reduces the occupation number of the oscillator by one: $a\ket{n}=\sqrt{n}\ket{n-1}$.
{[}$a=(x+i\frac{p}{m\omega})/(2x_{0})$ with $x_{0}=\sqrt{\hbar/(2m\omega)}$;
$x$ and $p$ being position and momentum operators for the oscillator]
There are many important applications: For example, a single standing
wave mode of the electromagnetic field inside an optical or microwave
cavity is described as a harmonic oscillator. That oscillator is damped
because the photons can leak out of the cavity through the semi-transparent
mirrors of the cavity, and the individual decay process corresponds
to the destruction of a single photon inside the cavity: $n\mapsto n-1$.
Another example of current interest are nanomechanical systems (small
beams on the micrometer scale) which are harmonic oscillators to a
good approximation, damped because of their connection to the mechanical
structure onto which they are attached and into which they can radiate
phonons.
\end{itemize}
We have not described how to obtain the rates and the operators, and
under which conditions one may expect the physical dynamics to be
well approximated by a Markov master equation. While a thorough discussion
of these points would go beyond the scope of the present notes, and
it is hard to list simple, generally valid conditions, the following
rule-of-thumb may be offered: most systems in which this approximation
works have decay rates that are both much smaller than the typical
transition frequencies of the system, and also smaller than the inverse
correlation time of the fluctuating force that the environment imposes
on the system. This condition is usually achieved in the limit of
a weak coupling between system and bath, as the decay and dephasing
rates become arbitrarily small in that limit.

\section{Beyond Lindblad equations}

Although in practice Lindblad Markov master equations are used in
the majority of applications in the various subfields of physics,
current research in quantum dissipative systems focuses on the interesting
effects that arise outside of this framework. Here we just list a
few of those physical situations and features, to give the reader
a sense of what lies beyond Lindblad dynamics. 

\begin{itemize}
\item In some areas it is experimentally feasible to measure the quanta
which are emitted into the environment by the system, thereby learning
more about the system's state. For example, the photons having been
emitted from an atom or leaking out of an optical cavity can be registered
by photo-detectors. The dynamics of the reduced density matrix, conditioned
on the observed detector results, may then often be described by a
modified master equation that depends on these results. Such an approach
sometimes is referred to as {}``quantum jump trajectories simulations'',
because it was initially developed to describe quantum jumps in individual
atoms that have been observed by light scattering.
\item Non-Markovian dynamics is generated when the environment's fluctuations
display long correlation times. A generic consequence of taking into
account these effects are short-time deviations from exponential decay.
In addition, under some circumstances the coherence of the system
can even be revived at later times, after having decayed initially.
Non-Markovian dynamics is generated when the environment's fluctuations
display long correlation times. A generic consequence of taking into
account these effects are short-time deviations from exponential decay.
In addition, under some circumstances the coherence of the system
can even be revived at later times, after having decayed initially.
Non-Markovian dynamics is generated when the environment's fluctuations
display long correlation times. A generic consequence of taking into
account these effects are short-time deviations from exponential decay.
In addition, under some circumstances the coherence of the system
can even be revived at later times, after having decayed initially.
\begin{figure}
\begin{centering}
\includegraphics[width=0.9\columnwidth]{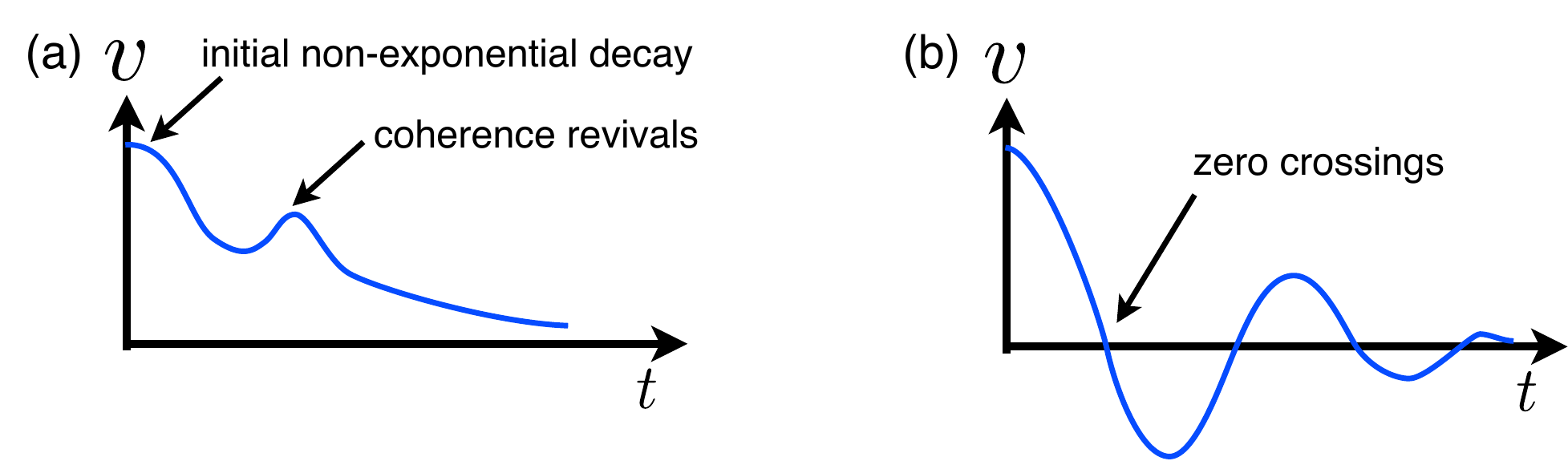}\caption{Possible features of the visibility in the example of pure dephasing
(section \ref{examplepuredeph}), due to non-Markovian dynamics (a)
and due to non-Gaussian noise (b). }

\par\end{centering}
\end{figure}

\item A favorite microscopic model for the environment is a bath of harmonic
oscillators, which is, for example, an exact representation of the
electromagnetic field or the phonons in a crystal lattice. When such
a bath is coupled to a single particle (bilinearly in the coordinates
of particle and bath oscillators), one speaks of the {}``Caldeira-Leggett
model'' \citep{1981_01_CaldeiraLeggett_TunnelingWithDissipation,1983_CaldeiraLeggett_QuantumBrownianMotion,2000_Weiss_QuantumDissipativeSystems}.
This allows to answer questions that go beyond Markov dynamics. For
example, when such a bath gives rise to standard diffusive motion
($\left\langle r^{2}(t)\right\rangle \propto t$) at high temperatures,
the {}``Quantum Brownian motion'' resulting at zero temperature
is sub-diffusive, with the distance to the origin obeying $\left\langle r^{2}(t)\right\rangle \propto\ln(t)$.
One can use the same kind of model to study the suppression of quantum
tunneling due to the influence of a dissipative environment. These
nonperturbative studies are usually carried out in a path-integral
framework.
\item As the coupling to the environment is increased, there can be qualitative
changes in behaviour at some critical coupling strength. The best
known example occurs in a model where a particle that tunnels between
two states is coupled to a bath of harmonic oscillators. This is commonly
referred to as the {}``spin-boson model'', where the {}``spin''
refers to the two-level system and {}``boson'' refers to the bath
of oscillators. For a certain distribution of bath oscillator frequencies
({}``Ohmic bath''), increasing the coupling strength beyond some
point makes the system undergo a quantum phase transition towards
a symmetry-broken phase where the particle remains trapped in one
of the two states for all times. Ergodicity is broken due to the strong
dissipation.
\item Even if the actual physical environment is not a bath of oscillators,
it often may be treated as such to a very good degree of approximation
as long as the coupling is weak. The fluctuating force generated by
a bath of oscillators is Gaussian-distributed. However, for strong
coupling, this approach will fail, and the dissipative dynamics can
become qualitatively different due to the non-Gaussian fluctuations
acting on the system. An example of current interest are the current
fluctuations generated by the passage of discrete, single electrons
through nanostructures. 
\end{itemize}

\section{Further reading}

Mathematical treatments of quantum dissipative dynamics may be found
in the books by Davies \citep{1976_Davies_BookDissipativeQM} (published
in 1976, still without reference to Lindblad, but developing the same
concepts) and Kraus \citep{1983_Kraus_StatesEffects}, who emphasizes
the connections with the theory of measurement. The treatise of von
Neumann on the mathematical foundations of quantum mechanics \citep{1932_vonNeumann_Book}
forms the basis.

A thorough introduction to the density matrix and its uses, at an
elementary level for the physicist, can be found in the book by Blum
\citep{1996_Blum_DensityMatrix}, which also presents a microscopic
derivation of the rates and relaxation operators appearing in master
equations. Dissipative quantum systems in general (often with emphasis
on quantum optics) are described in more detail in the books by Carmichael
\citep{1993_Carmichael_Book}, Gardiner and Zoller \citep{2004_GardinerZoller_QuantumNoise},
and Breuer and Petruccione \citep{2002_Breuer_Book}. The book by
Weiss \citep{2000_Weiss_QuantumDissipativeSystems} emphasizes those
aspects that go beyond the Lindblad master equations, such as the
Feynman-Vernon influence functional formalism, some exact solutions,
and a very detailed discussion of the spin-boson model. In that context,
the classic review by Leggett and co-workers on the spin-boson model
is also highly recommended \citep{1987_Leggett_ReviewSpinBoson}.

Kraus operators in the context of quantum information processing are
described in the book by Nielsen and Chuang \citep{2000_NielsenChuang_QuantumComputation},
as well as in the quantum information lecture notes by Preskill (\citep{1998_preskill_notes},
chapter 3), who also describes the Lindblad master equation and gives
a nice proof of the Kraus representation theorem. 

\bibliographystyle{plain}
\bibliography{/Users/florian/pre/bib/BibFM}

\begin{thebibliography}{10}

\bibitem{1996_Blum_DensityMatrix}
K.~Blum.
\newblock {\em {Density Matrix Theory and Applications}}.
\newblock Springer-Verlag, Berlin, 1996.

\bibitem{2002_Breuer_Book}
H.~P. Breuer and F.~Petruccione.
\newblock {\em The {Theory} of {Open Quantum Systems}}.
\newblock Oxford University Press (Oxford), 2002.

\bibitem{1981_01_CaldeiraLeggett_TunnelingWithDissipation}
A.~O. Caldeira and A.~J. Leggett.
\newblock Influence of dissipation on quantum tunneling in macroscopic systems.
\newblock {\em \prl}, 46:211, 1981.

\bibitem{1983_CaldeiraLeggett_QuantumBrownianMotion}
A.~O. Caldeira and A.~J. Leggett.
\newblock Path integral approach to quantum brownian motion.
\newblock {\em Physica}, 121A:587, 1983.

\bibitem{1993_Carmichael_Book}
H.~Carmichael.
\newblock {\em {An Open Systems Approach to Quantum Optics}}.
\newblock Springer-Verlag, Berlin, 1993.

\bibitem{1976_Davies_BookDissipativeQM}
E.~B. Davies.
\newblock {\em {Quantum Theory of Open Systems}}.
\newblock Academic Press, 1976.

\bibitem{1963_FeynmanVernon_InfluenceFunctional}
R.~P. Feynman and F.~L. Vernon.
\newblock The theory of a general quantum system interacting with a linear
  dissipative system.
\newblock {\em Annals of Physics (New York)}, 24:118, 1963.

\bibitem{2004_GardinerZoller_QuantumNoise}
C.~W. Gardiner and P.~Zoller.
\newblock {\em {Quantum Noise (Springer Verlag, Berlin)}}.
\newblock Springer-Verlag (Berlin), 2004.

\bibitem{1927_Heisenberg_UncertaintyRelation}
W.~Heisenberg.
\newblock {\em Zeitschrift {f\"ur} Physik}, 43:172, 1927.

\bibitem{1932_vonNeumann_Book}
{J. v. Neumann}.
\newblock {\em {Mathematical Foundations of Quantum Mechanics}}.
\newblock Princeton University Press, 1996.

\bibitem{1983_Kraus_StatesEffects}
K.~Kraus.
\newblock {\em States, Effects, and Operations Fundamental Notions of Quantum
  Theory: Lectures in Mathematical Physics at the University of Texas at
  Austin}, volume 190 of {\em Springer Lecture Notes in Physics}.
\newblock Springer-Verlag, Berlin, 1983.

\bibitem{1987_Leggett_ReviewSpinBoson}
A.~J. Leggett, S.~Chakravarty, A.~T. Dorsey, M.~P.~A. Fisher, A.~Garg, and
  W.~Zwerger.
\newblock Dynamics of the dissipative two-state system.
\newblock {\em Rev. Mod. Phys.}, 59:1, 1987.

\bibitem{1976_Lindblad_MasterEquation}
G.~Lindblad.
\newblock On the generators of quantum dynamical semigroups.
\newblock {\em Commun. math. Phys.}, 48:119--130, 1976.

\bibitem{2000_NielsenChuang_QuantumComputation}
M.~A. Nielsen and I.~L. Chuang.
\newblock {\em Quantum computation and quantum information}.
\newblock Cambridge University Press, Cambridge, 2000.

\bibitem{1996_Peres_SeparabilityCriterion}
A.~Peres.
\newblock Separability criterion for density matrices.
\newblock {\em \prl}, 77:1413, 1996.

\bibitem{1998_preskill_notes}
J.~Preskill.
\newblock Lecture notes for physics 229: Quantum information and computation.
\newblock Technical report, California Institute of Technology, 1998.

\bibitem{1972_ReedSimon_FunctionalAnalysisBook}
M.~Reed and B.~Simon.
\newblock {\em Methods of modern mathematical physics. I. Functional analysis}.
\newblock Academic Press, 1972.

\bibitem{2000_Weiss_QuantumDissipativeSystems}
U.~Weiss.
\newblock {\em Quantum Dissipative Systems}.
\newblock World Scientific, Singapore, 2000.

\end{thebibliography}

\end{document}